

\input phyzzx.tex
\overfullrule0pt

\def\ie{{\it i.e.}}
\def\etal{{\it et al.}}
\def\half{{\textstyle{1 \over 2}}}

\def\bold#1{\setbox0=\hbox{$#1$}%
     \kern-.025em\copy0\kern-\wd0
     \kern.05em\copy0\kern-\wd0
     \kern-.025em\raise.0433em\box0 }
\Pubnum={VAND-TH-94-13}
\date={August 1994}
\pubtype{}
\titlepage


\vskip1cm
\title{\bf Is a  Light Gluino Compatible with N=1 Supersymmetry?${}^{*}$
}
\vskip .1in
\author{Marco Aurelio D\'\i az }
\vskip .1in
\centerline{Department of Physics and Astronomy}
\centerline{Vanderbilt University, Nashville, TN 37235${}^{\#}$}
\vskip .2in

\centerline{\bf Abstract}
\vskip .1in

Using the new CLEO bound on the branching ratio of the decay mode
$b\longrightarrow s\gamma$ given by $1\times10^{-4}<
B(b\longrightarrow s\gamma)<4\times10^{-4}$ at 95\% c.l., and
the experimental bounds on the masses of the lightest chargino,
the second lightest neutralino, and the light CP-odd Higgs, we find
that a light gluino is incompatible with a class of supergravity
models defined by $N=1$ supergravity, with a radiatively broken
electroweak symmetry group and universality of scalar and gaugino
masses at the unification scale. We also find in this scenario
strong constraints on the parameter space due to the new CLEO
bounds on $B(b\longrightarrow s\gamma)$.

\vskip 2cm
\noindent * Presented in the International Workshop ``SUSY-94'',
University of Michigan, Ann Arbor MI, May 14-17, 1994.

\noindent ${}^{\#}$ Address starting Sept. 1, 1994: Physics Department,
University of Southampton, Southampton SO9 5NH, United Kingdom.

\vfill

\endpage

\voffset=-0.2cm

\REF\gaugeunif{M.B. Einhorn and D.T.R. Jones, {\it Nucl. Phys.}
{\bf B196}, 475 (1982); W.J. Marciano and G. Senjanovic, {\it
Phys. Rev. D} {\bf 25}, 3092 (1982); U. Amaldi \etal, {\it Phys.
Rev. D} {\bf 36}, 1385 (1987); U. Amaldi
, W. de Boer, and H. Furstenau,
{\it Phys. Lett. B} {\bf 260}, 447 (1991); U. Amaldi
\etal, {\it Phys. Lett.} {\bf B281}, 374 (1992).}

It is well known that in the Standard Model (SM)
the three gauge couplings $g_s$, $g$, and $g'$,
corresponding to the gauge groups $SU(3)\times SU(2)\times U(1)$,
do not converge to a single value when we run these couplings up to
scales near the Planck scale. Although it is not a proof of
supersymmetry, it is interesting that within the minimal
supersymmetric extension of the Standard Model (MSSM) this gauge
coupling unification can be achieved\refmark\gaugeunif.

\REF\softterm{L. Girardello and M.T. Grisaru, {\it Nucl. Phys.}
{\bf B194}, 65 (1982).}

In supersymmetry, fermionic and bosonic degrees of freedom are related
by a symmetry. If the symmetry is unbroken, every known fermion
(boson) has a bosonic (fermionic) supersymmetric partner degenerate in
mass. Differences in mass appear between partners as soon as
supersymmetry is broken. This is achieve through soft-supersymmetry
terms which do not introduce quadratic divergences to the
unrenormalized theory\refmark\softterm.

\REF\lightold{G.R. Farrar and P. Fayet, {\it Phys. Lett.} {\bf
76B}, 575 (1978); T. Goldman, {\it Phys. Lett.} {\bf 78B}, 110
(1978); I. Antoniadis
, C. Kounnas, and R. Lacaze,
{\it Nucl. Phys.} {\bf B211}, 216 (1983); M.J. Eides and M.J. Vysotsky,
{\it Phys. Lett.} {\bf 124B}, 83 (1983); E. Franco, {\it Phys.
Lett.} {\bf 124B}, 271 (1983); G.R. Farrar, {\it Phys. Rev.
Lett.} {\bf 53}, 1029 (1984);
J. Ellis and H. Kowalski, {\it Phys. Lett. B} {\bf
157}, 437 (1985); V. Barger
, S. Jacobs, J. Woodside, and K. Hagiwara,
{\it Phys. Rev. D} {\bf 33}, 57 (1986).}
\REF\lightnew{I. Antoniadis
, J. Ellis, and D.V. Nanopoulos,
{\it Phys. Lett. B} {\bf 262}, 109 (1991); L. Clavelli, {\it Phys.
Rev. D} {\bf 46}, 2112 (1992); M. Jezabek and J.H. K\"uhn, {\it
Phys. Lett. B} {\bf 301}, 121 (1993); J. Ellis
, D.V. Nanopoulos, and D.A. Ross,
{\it Phys. Lett. B} {\bf 305}, 375 (1993);
R.G. Roberts and W.J. Stirling, {\it Phys. Lett. B} { \bf 313},
453 (1993); M. Carena
, L. Clavelli, D. Matalliotakis, H.P. Nilles, and C.E.M. Wagner,
{\it Phys. Lett. B} {\bf 317}, 346 (1993);
F. Cuypers, {\it Phys. Rev. D} {\bf 49}, 3075 (1994);
R. Mu\~noz-Tapia and W.J. Stirling, {\sl Phys. Rev. D}
{\bf 49}, 3763 (1994);
D.V. Shirkov and S.V. Mikhailov, Report No. BI-TP 93/75,
Jan. 1994.}
\REF\CValle{F. de Campos and J.W.F. Valle, Report No. FTUV/93-9,
Feb. 1993.}
\REF\LNW{J.L. Lopez
, D.V. Nanopoulos, and X. Wang,
{\it Phys. Lett. B} {\bf 313}, 241 (1993).}

The supersymmetric partner
of the gluon is the gluino, and discussions about the existence of
a light gluino have been in the literature for some
time\refmark\lightold.
Motivated by the discrepancy between the value of the strong coupling
constant $\alpha_s$, determined by low energy deep inelastic
lepton-nucleon scattering: $\alpha_s(m_Z)=0.112\pm0.004$, and the one
determined by high energy $e^+e^-$ LEP experiments: $\alpha_s=0.124
\pm0.005$, there has been a renewed interest in the possibility of a
light gluino\refmark{\lightnew-\LNW}.

\REF\CUSB{P.M. Tuts \etal, {\sl Phys. Lett. B} {\bf 186}, 233 (1987).}
\REF\UAuno{UA1 Collaboration, C. Albajar \etal, {\sl Phys. Lett. B}
{\bf 198}, 261 (1987).}
\REF\howie{H.E. Haber, Report No. SCIPP 93/21, July 1993;
R.M. Barnett
, H.E. Haber, and G.L. Kane,
{\it Nucl. Phys.} {\bf B267}, 625 (1986).}
\REF\CFarrar{M.B. \c Cak\i r and G.R. Farrar, Report Number
RU-94-04, January 1994.}

An analysis of $\Upsilon$ decays in the CUSB detector excluded
gluinos with a mass $0.6<m_{\tilde g}<2.2$ GeV\refmark\CUSB.
Combining this and other results the UA1 Collaboration found
that the gluino mass is allowed in the region $2.6<m_{\tilde g}\lsim
6$ GeV and $m_{\tilde g}<0.6$ GeV\refmark\UAuno. These bounds are
controversial and, according to ref.~[\howie], the upper bound is
$m_{\tilde g}\lsim 3$ GeV. Even more, recently ref.~[\CFarrar]
pointed out that results from quarkonium decays cannot rule out
gluino masses below about 2 GeV [$m(\eta_{\tilde g})\lsim 3$ GeV,
where $\eta_{\tilde g}$ is a pseudoscalar $\tilde g\tilde g$
bound state].

\REF\susyrep{H.P. Nilles, {\it Phys. Rep.} {\bf 110}, 1 (1984);
H.E. Haber and G.L. Kane, {\it Phys. Rep.} {\bf 117}, 75 (1985);
R. Barbieri, {\it Riv. Nuovo Cimento} {\bf 11}, 1 (1988).}

One of the most successful supersymmetric models is minimal $N=1$
supergravity, in which the electroweak symmetry breaking can be
achieved radiatively\refmark\susyrep through the evolution of the Higgs
mass parameters from the unification scale to the weak scale.
In this model, the three
gaugino masses $M_s$, $M$, and $M'$ are different at the weak scale
but equal to a common gaugino mass $M_{1/2}$ at the grand unification
scale $M_X$. The difference at the weak scale is due to the fact that
the evolution of the three masses is controlled by different
renormalization group equations (RGE). The approximated solution of
these RGE is:
$$\eqalign{M_s&\approx M_{1/2}\left[1+{{3g_s^2}\over
{8\pi^2}}\ln{{M_X}\over{m_Z}}\right],\qquad m_{\tilde g}=|M_s|
\cr M&\approx M_{1/2}\left[1-{{g^2}\over{8\pi^2}}\ln{{M_X}\over{m_Z}}
\right]
\cr M'&\approx M_{1/2}\left[1-{{11{g'}^2}\over{8\pi^2}}\ln{{M_X}\over{
m_Z}}\right]\cr}\eqn\RGEsol$$
where we are neglecting the supersymmetric
threshold effects. Taking $M_X=10^{16}$ GeV, we find that
$M\approx 0.30\,m_{\tilde g}$
and $M'\approx 0.16\,m_{\tilde g}$.

Similarly,
the scalar masses are also degenerate at the unification scale, and
equal to $m_0$. The RGE make both the Higgs mass parameters
$m_1$ and $m_2$, and the squark and slepton mass parameters,
evolve differently.
A third independent parameter at the unification scale
is the mass parameter $B$. This mass defines the value of the unified
trilinear mass parameter $A$ at $M_X$ by $A=B+m_0$, a
relation valid in models with canonical kinetic terms. Moreover, it
also defines the
third Higgs mass parameter $m_{12}^2=-B\mu$, valid at every scale,
where $\mu$ is the supersymmetric Higgs mass parameter.
The set of independent parameters we choose to work with, given by
$M_{1/2}$, $m_0$, and $B$ at the unification scale,
is completed by the value of the top quark Yukawa coupling
$h_t=gm_t/(\sqrt{2}m_Ws_{\beta})$
at the weak scale. Here the angle $\beta$ is defined through
$\tan\beta=v_2/v_1$, where $v_1$ and $v_2$ are the vacuum expectation
values of the two Higgs doublets. We define the top Yukawa coupling
in a on-shell scheme.

\REF\neutralinorad{A.B. Lahanas
, K. Tamvakis, and N.D. Tracas,
{\it Phys. Lett. B} {\bf 324}, 387 (1994);
D. Pierce and A. Papadopoulos, {\sl Phys. Rev. D} {\bf 50}, 565 (1994),
and JHU-TIPAC-940001, Feb. 1994.}

Knowing the parameters of the Higgs potential at the weak scale
$m_1^2$, $m_2^2$, and $B$, we can calculate the more familiar
parameters $m_t$, $t_{\beta}$, $m_A$, and $\mu$, for a given
value of the top quark Yukawa coupling $h_t$, through the
following formulas valid at tree level
$$\eqalign{m_{1H}^2\equiv m_1^2+\mu^2=&-\half m_Z^2c_{2\beta}+\half
m_A^2(1-c_{2\beta}),
\cr m_{2H}^2\equiv m_2^2+\mu^2=&\half m_Z^2c_{2\beta}+\half
m_A^2(1+c_{2\beta}),\cr
m_{12}^2=-B\mu=&\half m_A^2s_{2\beta},\cr}\eqn\conversion$$
where $s_{2\beta}$ and $c_{2\beta}$ are sine and cosine functions of
the angle $2\beta$, and it is understood that all the parameters
are evaluated at the weak scale. We alert the reader that for a
given set of values $M_{1/2}$, $m_0$, $B$, and $h_t$ there may
exist more than one solution for the parameters at the weak scale
$m_t$, $t_{\beta}$, $m_A$, and $\mu$. According to ref.~[\LNW],
and we will confirm this, the relevant region of parameter space
in the light gluino scenario is characterized by low values of the
top quark mass and values of $\tan\beta$ close to unity.
Considering the low values of the top
quark mass relevant for our calculations,
radiative corrections to the chargino and
neutralino masses (recently calculated in ref.~[\neutralinorad])
will have a minor effect.

\REF\mbmtau{M. Carena
, S. Pokorski, and C.E.M. Wagner,
{\it Nucl. Phys.} {\bf B406}, 59 (1993); P. Langacker and N. Polonsky,
{\it Phys. Rev. D} {\bf 49}, 1454 (1994).}
\REF\AnantBS{B. Ananthanarayan
, K.S. Babu, and Q. Shafi,
Report No. BA-94-02, Jan. 1994.}
\REF\diazhaberii{M.A. D\'\i az and H.E. Haber, {\it Phys. Rev. D}
{\bf 46}, 3086 (1992).}
\REF\aleph{D. Buskulic \etal, {\it Phys. Lett. B} {\bf 313}, 312
(1993).}

The region $\tan\beta$ close to unity has been singled out by the
grand unification condition $m_b=m_{\tau}$ at $M_X$\refmark\mbmtau,
and was analyzed in detail in ref.~[\AnantBS]. Here we do not impose
the Yukawa unification, but we stress the
fact that if $\tan\beta=1$, the lightest CP-even neutral Higgs is
massless at tree level. Nevertheless, the supersymmetric
Coleman-Weinberg mechanism\refmark\diazhaberii generates a mass $m_h$
different from zero via radiative corrections. The fact that
$m_t$ is also small will result in a radiatively generated $m_h$
close to the experimental lower limit $m_h\gsim56$ GeV, valid for
$m_A>100$ GeV\refmark\aleph. Therefore, experimental lower limits
on $m_h$ impose important restrictions on the light gluino window.

\REF\BBMR{S. Bertolini
, F. Borzumati, A. Masiero, and G. Ridolfi,
{\it Nucl. Phys.} {\bf B353}, 591 (1991).}
\REF\masSUSY{M.A. D\'\i az, {\it Phys. Lett. B} {\bf 304},
278 (1993); N. Oshimo, {\it Nucl. Phys.} {\bf B404}, 20 (1993);
J.L. Lopez
, D.V. Nanopoulos, and G.T. Park,
{\it Phys. Rev. D} {\bf 48}, 974 (1993);
Y. Okada, {\it Phys. Lett. B} {\bf 315}, 119 (1993);
R. Garisto and J.N. Ng, {\it Phys. Lett. B} {\bf 315}, 372 (1993);
J.L. Lopez
, D.V. Nanopoulos, G.T. Park, and A. Zichichi,
{\it Phys. Rev. D} {\bf 49}, 355 (1994);
M.A. D\'\i az, {\it Phys. Lett. B} {\bf 322}, 207 (1994);
F.M. Borzumati, Report No. DESY 93-090, August 1993; S. Bertolini
and F. Vissani, Report No. SISSA 40/94/EP, March 1994.}
\REF\cleo{CLEO Collaboration, B. Barish \etal, Report Number
CLEO CONF 94-1.}
\REF\misiak{M. Misiak, {\it Nucl. Phys.} {\bf B393}, 23 (1993).}
\REF\ChargedH{J.F. Gunion and A. Turski, {\it Phys. Rev. D} {\bf 39},
2701 (1989); {\bf 40}, 2333 (1989); A. Brignole
, J. Ellis, G. Ridolfi, and F. Zwirner,
{\it Phys. Lett. B} {\bf 271}, 123 (1991); M. Drees
and M.M. Nojiri, {\it Phys. Rev. D} {\bf 45}, 2482 (1992);
A. Brignole, {\it Phys. Lett. B} {\bf 277}, 313 (1992);
P.H. Chankowski
, S. Pokorski, and J. Rosiek,
{\it Phys. Lett. B} {\bf 274}, 191 (1992);
M.A. D\'\i az and H.E. Haber, {\it Phys. Rev. D}
{\bf 45}, 4246 (1992).}
\REF\diaz{M.A. D\'\i az, {\it Phys. Rev. D} {\bf 48}, 2152 (1993).}

It has been pointed out that the branching ratio
$B(b\longrightarrow s\gamma)$ has a strong dependence on the
supersymmetric parameters\refmark{\BBMR,\masSUSY}.
The theoretical branching ratio must remain within the new\refmark\cleo
experimental bounds $1\times10^{-4}<B(b\longrightarrow s\gamma)<
4\times10^{-4}$ at 95\% c.l.
We calculate this ratio, including loops involving
$W^{\pm}$/U-quarks, $H^{\pm}$/U-quarks, $\chi^{\pm}$/U-squarks,
and $\tilde g$/D-squarks, neglecting only the contribution from
the neutralinos, which were
reported to be small\refmark\BBMR. We also include
QCD corrections to the branching ratio\refmark\misiak and
one loop electroweak corrections to both the charged Higgs
mass\refmark\ChargedH and the charged Higgs-fermion-fermion
vertex\refmark\diaz.

\REF\neuDelphi{G. Wormser, published in {\it Dallas HEP 1992},
page 1309.}
\REF\Decamprep{D. Decamp \etal (ALEPH Collaboration), {\it Phys. Rep.}
{\bf 216}, 253 (1992).}

Another important source of constraints comes from the
chargino/neutralino sector. For $\tan\beta\gsim4$, a neutralino with
mass lower than 27 GeV is excluded, but the lower bound decreases
when $\tan\beta$ decreases, and no bound is obtained if
$\tan\beta<1.6$\refmark\neuDelphi. The lower bound for the heavier
neutralinos (collectively denoted by $\chi'$) is $m_{\chi'}>45$ GeV
for $\tan\beta\gsim 3$, and this bound also decreases with $\tan\beta$
and eventually disappears\refmark\Decamprep.
On the other hand, if the
lightest neutralino has a mass $\lsim 40$ GeV (as we will see,
in the light gluino scenario, the lightest neutralino has a mass
of the order of 1 GeV), the lower bound
for the lightest chargino mass is 47 GeV\refmark\Decamprep. For
notational convenience, this latest experimental bound will be denoted
by $\bar m_{\chi^{\pm}_1}\equiv47$ GeV.

\REF\diazlglu{M.A. D\'\i az, Report No. VAND-TH-94-7, revised
version, March 1994.}

In the light gluino scenario we have $M_{1/2}\approx0$, in this way
the chargino masses depend only on two parameters: $\mu$ and
$\tan\beta$. The experimental lower bound on $m_{\chi^{\pm}_1}$
limits the values of $\mu$ and $\tan\beta$\refmark\diazlglu:
$$\eqalign{
\mu^2&<\bar m_{\chi^{\pm}_1}^2\left({{m_W^2}\over{\bar
m_{\chi^{\pm}_1}^2}}-1\right)^2-{{4m_W^2\mu_0M(\half\mu_0^2+m_W^2-
\bar m_{\chi^{\pm}_1}^2)}\over
{\bar m_{\chi^{\pm}_1}^2|\mu_0|\sqrt{\mu_0^2+4m_W^2}}}
+{\it O}(M_{1/2}^2)\cr
&\qquad\qquad\Longrightarrow|\mu|\lsim(90\mp0.87m_{\tilde g})\,
{\rm GeV},\,\,{\rm with}\qquad \pm={\rm sign}(\mu M)\cr
|c_{2\beta}|&<1-{{\bar m_{\chi^{\pm}_1}^2}\over{m_W^2}}+{\cal O}
(M_{1/2}^2)\quad\Longrightarrow\quad
0.46<t_{\beta}<2.2\cr}\eqn\constrain$$
where $\mu_0^2=\bar m^2_{\chi^{\pm}_1}(m_W^2/\bar m_{\chi^{\pm}_1}^2
-1)^2\approx90$ GeV is the zero order solution ($M=0$), and
$\mp0.87m_{\tilde g}$ correspond to the first order correction.
The type of constraints given
in eq.~\constrain\ were already found in
ref.~[\LNW] at zero order, but as we will see, the neutralino sector
will restrict the parameter space even more.

\REF\GunionHaber{J.F. Gunion, H.E. Haber, {\it Nucl. Phys.} {\bf B272},
1 (1986).}

The neutralino mass matrix in the zero gluino mass limit ($M=M'=0$)
has one eigenvalue equal to zero. Including
first order corrections, we get for the lightest
neutralino mass a value close to 0.5 GeV.
This light neutralino (the lightest supersymmetric particle, or LSP)
is, up to terms of ${\cal O}(M_{1/2}^2/m_Z^2)$, almost a pure photino,
and there is no bound on its mass from LEP collider data. Nevertheless,
in the case of a stable LSP (R-parity conserving models), ref.~[\LNW]
pointed out some cosmological implications that make this scenario
less attractive. On the other hand, the possibility of having a
small amount of R-parity violation is not ruled out, in which case
the LSP would not be stable\refmark\CValle.
The second lightest neutralino is a higgsino with a mass close
to the absolute value of $\mu$, and experimental bounds on its mass
will impose important restrictions on the model.

\FIG\figi{Contours of constant value of the lightest chargino and
the second lightest neutralino
masses, for a gluino mass $m_{\tilde g}=3$ GeV. The contour
corresponding to the chargino mass is
defined by the experimental lower bound
$m_{\chi^{\pm}_1}=47$. For $\chi^0_2$ we plot contour of constant mass
from 5 to 45 GeV (dashed lines). The solid line that joins the
crosses represent the $\tan\beta$ dependent bound on $m_{\chi^0_2}$.
The ``allowed'' region lies below the two solid lines
(chargino/neutralino searches only).}

Now we turn to the exact numerical calculation of
the chargino and neutralino masses. In Fig.~\figi\ we plot contours
of constant masses in the $\mu-t_{\beta}$ plane. The
curve $m_{\chi^{\pm}_1}=47$ GeV corresponds to the experimental
constraint on $m_{\chi^{\pm}_1}$. We also plot contours
defined by $m_{\chi^0_2}=5-45$ GeV, and the $\tan\beta$ dependent
experimental bound on $m_{\chi^0_2}$ is represented by the solid line
that joins the crosses. In this way, the ``allowed'' region
(including chargino/neutralino searches only) corresponds to the region
below the two solid lines. For $\mu<0$ the allowed region
is almost an exact reflection. The approximate bounds
for $\mu$ we got in eq.~\constrain\ are confirmed numerically:
$\mu<87.4$ GeV for $m_{\tilde g}=3$ GeV. Nevertheless, the bounds
on $\tan\beta$ come only from the experimental
result $m_{\chi^{\pm}_2}>47$ GeV, and we must include also the
experimental results on $m_{\chi^0_2}$. From Fig.~\figi\ we see
that this bound restricts the model to
$\tan\beta\lsim1.82$, with the equality valid for
$\mu=49.4$ GeV. Since for $\tan\beta\lsim1$
there is no solution for the radiatively broken
electroweak symmetry group, the allowed values
of $\tan\beta$ in the light gluino scenario and with $\mu>0$ are
$$1\lsim\tan\beta\lsim1.82\,.\eqn\tanballow$$
If $\mu<0$, the upper bound is $\tan\beta\lsim 1.85$ with the equality
valid for $\mu=-51.8$ GeV. We go on to analyze the
viability of the ``allowed'' region in Fig.~\figi. We will find that
the region allowed by the $\chi^{\pm}$ and $\chi^0$ analysis is in fact
disallowed by the experimental bound on $m_h$ and $m_t$.

\REF\solRGE{R. Barbieri and G.F. Giudice, {\it Nucl. Phys.}
{\bf B306}, 63 (1988); M. Matsumoto
, J. Arafune, H. Tanaka, and K. Shiraichi,
{\it Phys. Rev. D} {\bf 46}, 3966 (1992).}
\REF\topbound{S. Abachi \etal (D0 Collaboration), {\it Phys. Rev.
Lett.} {\bf 72}, 2138 (1994).}

In ref.~[\solRGE]\ the RGE are solved for the special case in which
only the top quark Yukawa coupling is different from zero. In the
case of a light gluino ($M_{1/2}\approx0$),
the value of $\mu$ at the weak scale can be approximated
by\refmark\solRGE
$$\half m_Z^2+\mu^2=-m_0^2+{{z-1}\over{z(1-t_{\beta}^{-2})}}
\left[{{3m_0^2}\over2}+{{A^2}\over{2z}}\right],\eqn\solmu$$
with
$$z^{-1}=1-(1+t_{\beta}^{-2})\left({{m_t}\over{193 {\rm GeV}}}
\right)^2.\eqn\zeta$$
As it was reported in ref.~[\LNW], there is a fine-tuning situation
in which we can have $m_0\gg|\mu|$ (producing larger radiative
corrections to $m_h$) and it is obtained when the
coefficient of $m_0^2$ in eq.~\solmu\ is zero. Ref.~[\LNW]\ concluded
that constraints on $m_h$ can be satisfied in a small window
around $\tan\beta=1.88-1.89$ (they did not consider the constraint
on the second lightest neutralino). We will see that if the
relation $A=B+m_0$ holds we do not find this type of solution
($m_0\gg|\mu|$) as opposed to the case in which
$A=0$\refmark\diazlglu. However, the
later is obtained for a value of the top quark mass below the
value of the experimental lower bound $m_t\ge131$ GeV
\refmark\topbound.

We survey the parameter space $m_0$, $B$, $M_{1/2}$, and $h_t$,
looking for the maximum value
of $\tan\beta$ allowed by collider negative searches in the
chargino/neutralino sector, using the SUSY-GUT model described earlier.
We consider first models in which the relation $A=B+m_0$ holds.
We expect maximum $\tan\beta$ to maximize $m_h$. For example,
for the value $h_t=0.87$ and $M_{1/2}=1$ GeV
(essentially fixed by the light gluino mass hypothesis)
we find that $m_0=132.9$
and $B=-225.5$ GeV (at the unification scale) gives us
$\tan\beta=1.82$ and $\mu=49.4$ GeV, \ie, the
critical point with maximum $\tan\beta$ in the upper
corner of the allowed region in Fig.~\figi. The values of other
important parameters at the weak scale are
$m_{\chi^{\pm}_1}\approx47.1$, $m_{\chi^0_2}\approx36.8$,
$m_t=131.1$, $m_A=152.1$, and $m_{\tilde g}=2.75$ GeV. We find a
value for $B(b\longrightarrow s\gamma)=5.35\times10^{-4}$ inconsistent
with the CLEO bounds and the lightest CP-even neutral Higgs
also fails to meet the experimental
requirement: we get $m_h=47.7$ GeV, inconsistent with LEP data
since it is required that
$m_h>56$ GeV for $m_A>100$ GeV\refmark\aleph. It is known
that relative to the SM coupling, the $ZZh$ coupling in the MSSM
is suppressed by $\sin(\beta-\alpha)$, where $\alpha$ is
the mixing angle in the CP-even neutral Higgs sector, nevertheless,
this angle approaches the asymptotic value $\beta-\pi/2$ when
$m_A$ increases, \ie, the lightest Higgs $h$ behaves like the SM
Higgs and the experimental lower bound on its mass
will approach the SM bound.

\FIG\figii{For a fixed value of $M_{1/2}=1$ GeV and $h_t=0.87$ we
vary $m_0$: (a) $B=-225.5$ GeV and $m_0=61-151$ GeV; (b) $B=-200$
GeV and $m_0=51-133$ GeV (solid lines).
{}From chargino/neutralino searches only,
the experimentally allowed region lies below the two dashed lines.
In case (a) the curve passes through the
critical point defined by $\tan\beta=1.82$ and $\mu=49.4$ GeV.
In case (b), with a smaller value of the magnitude of $B$,
the curve passes below the critical point.}
\FIG\figiii{Masses of the lightest chargino (upper dashed line),
the second lightest neutralino (lower dashed line)
and the lightest CP-even Higgs (solid line) as a function of $\mu$
for the two cases in the previous figure: (a) $B=-225.5$ GeV and
(b) $B=-200$ GeV. The two horizontal doted lines correspond
to the experimental bounds $m_h>56$ GeV and $m_{\chi^{\pm}_1}>47$
GeV. The doted line joining the crosses represents the experimental
bound on the second lightest neutralino.}

In Fig.~\figii\ we take the critical value $B=-225.5$ GeV and vary
$m_0$ from 61 to 151 GeV [solid curve (a)] and we also take
$B=-200$ GeV and vary $m_0$ from 51 to 133 GeV [solid curve (b)].
The two dashed lines correspond to the experimental constraint
on the lightest chargino and the second lightest neutralino. The
``allowed'' region lies below both dashed curves.
Curve (a) intersect the ``allowed'' region in almost a point
(the critical point), as opposed to curve (b) which pass through
the ``allowed'' region below the critical point. We see that
low values of $m_0$ produce a too light neutralino and, on the
other hand, larger values of $m_0$ produce a too light chargino.
In Fig.~\figiii\ we can see the evolution of the
masses $m_h$, $m_{\chi^{\pm}_1}$, and $m_{\chi^0_2}$ in both cases.
Experimental bounds on $m_h$ and $m_{\chi^{\pm}_1}$ are represented
by horizontal doted lines, and the doted line joining the crosses
represent the experimental bound on the neutralino. In
Fig.~\figiii (a) we see that the bounds on chargino and neutralino
masses are satisfied only at the critical point but the lightest
CP-even Higgs mass is in conflict with its experimental bound in the
hole range. In Fig.~\figiii (b) the bounds on the chargino and
neutralino masses are satisfied in a wider region around the
critical point, however, the Higgs mass is even lighter than the
previous case.

\FIG\figiv{Masses of the top quark, the CP-odd neutral Higgs, and
the lightest up-type squark (mainly stop) at the weak scale, and
of the universal scalar mass at the unification scale as a function
of $\mu$ for the two cases defined in Fig.~\figii: (a) $B=-225.5$
GeV and (b) $B=-200$ GeV.}
\FIG\figv{Branching ratio $B(b\rightarrow s\gamma)$ as a function
of $\mu$ for the two cases (a) $B=-225.5$ GeV and (b) $B=-200$ GeV.}

In Fig.~\figiv\ we plot the evolution of the masses of the top quark
$m_t$, the CP-odd Higgs $m_A$, and the lightest up-type squark
$m_{\tilde q_{U1}}$ (mainly stop) at the weak scale, and of the
universal scalar mass parameter $m_0$ at the unification scale.
Similar behavior is observed in the two cases (a) and (b) related
to the two solid curves in Fig.~\figii. Larger values of $\mu$
are found when the value of $m_0$ is increased; similarly, $m_A$ and
$m_{\tilde q_{U1}}$ follow the same tendency. The top squark mass
starts to deviate from $m_0$ at larger values of $\mu$ due to the
increasing left-right squark mixing.

The prediction of the
branching ratio $B(b\rightarrow s\gamma)$ is plotted in Fig.~\figv\
for the two cases (a) and (b) described before. We see strong constraints
on the model, in fact, for the
particular case we are analysing here, it is evident  from the figure
an upper bound on the parameter $\mu$ given by $\sim 32$ GeV.
It will be interesting to see if this kind of bounds appear also
in different regions of parameters space (for example if $M_{1/2}\gg 1$
GeV). The branching ratio goes to zero for low values of the parameter
$\mu$ because the chargino contribution is large and negative
in that region.

{}From the two fixed parameters, $h_t$ and $M_{1/2}$, the one that could
affect the mass of the CP-even neutral Higgs is the first one; for a
fixed value of $\tan\beta$, a larger value of the top quark Yukawa
coupling will give us a larger $m_t$, and this will increase $m_h$.
However, $h_t$ also enters the RGE for the Higgs mass parameters,
and in order to get the correct electroweak symmetry breaking, a
smaller value of $m_0$ is necessary. This implies smaller squark
masses, which in turn reduce $m_h$ through radiative corrections.
As an example with a larger $h_t$, we
have found that for $h_t=0.97$ and $M_{1/2}=1$ GeV, the critical
point is obtained at $m_0=103.8$ and $B=-132.5$ GeV. As expected, the
value of the top quark mass is larger ($m_t=146.2$ GeV),
but we get smaller values for the squark masses.
The net effect is that now $m_h$ is even smaller, 43.5
GeV, also in conflict with the experimental lower bound. (We caution
the reader that at the small values of $m_t$ and $m_0$ used here,
the contributions to $m_h$ coming from the
Higgs/Gauge-boson/neutralino/chargino are also
important\refmark\diazhaberii; we include these in our analysis).
Also in conflict with experiment is the branching ratio
$B(b\longrightarrow s\gamma)=5.47\times10^{-4}$.

We go back to $h_t=0.87$ to analyze the case
$\mu<0$. In this case the critical point, given by $\tan\beta=1.85$
and $\mu=-51.8$ GeV, is obtained for $m_0=71.1$ and
$B=111$ GeV. However the light CP-even Higgs is lighter than before:
$m_h=40.4$ GeV, incompatible with LEP data, and
$B(b\longrightarrow s\gamma)=5.35\times10^{-4}$, incompatible with
CLEO bounds.

Models in which $A$ and $B$ are independent parameters have one
extra degree of freedom that may help to satisfy the experimental
constraints. According to eq.~\solmu\ the fine tuning $m_0\gg|\mu|$
is obtained for $A=0$. Adopting that value and considering $\mu>0$,
for $h_t=0.87$ and $M_{1/2}=1$
GeV we obtain the critical point for $m_0=151.6$ and $B=-256.8$ GeV,
which implies $m_{\chi^{\pm}_1}=47.1$, $m_{\chi^0_2}=36.8$,
$m_t=131.1$, $m_A=173.4$, and $m_{\tilde g}=2.75$ GeV. However,
we get $B(b\longrightarrow s\gamma)=7.15\times10^{-4}$ and
$m_h=48.8$ GeV, both inconsistent with experimental bounds. The
branching ratio $B(b\longrightarrow s\gamma)$ is larger than the
previous cases because the lightest up-type squark mass is larger,
decreasing the (negative) contribution of the up-squark/chargino
loops.

In order to illustrate the fine-tuning we consider $h_t=0.77$ and
$M_{1/2}=1$ GeV. The critical point is obtained for $m_0=930$
and $B=-9576$ GeV. The masses of $\chi^{\pm}_1$, $\chi^0_2$
and $\tilde g$ are the same as before, and we also get $m_A=1059$,
$m_h=64.8$ GeV consistent with LEP bound,
and $B(b\longrightarrow s\gamma)=4.14\times10^{-4}$
marginally consistent
with the CLEO bound, but this time it is the top quark mass
that does not meet the experimental bound: we get $m_t=116.0$ GeV,
incompatible with the $D0$ lower bound of 131 GeV. The branching
ratio $B(b\longrightarrow s\gamma)$ is smaller than before because
this time all the non-SM particles are heavy, in which case the
charged-Higgs/top-quark loop is also small and the ratio approach
to the SM value.

If $\mu<0$ no big changes are found. For $h_t=0.87$ the
critical point, defined now
by $\tan\beta=1.85$ and $\mu=-51.8$ GeV, is obtained for $m_0=159.4$
and $B=268$ GeV and, as before, the two quantities inconsistent
with experimental results are $B(b\longrightarrow s\gamma)=
8.42\times10^{-4}$ and $m_h=50.5$ GeV.

\REF\pomnir{N. Polonsky and A. Pomarol, Report Number PRINT-94-0149,
May 1994, and these proceedings.}

Our conclusion is that N=1 Supergravity with a radiatively broken
electroweak symmetry group is incompatible with a light gluino with
a mass of a few GeV. This is valid
in models where the relation $A=B+m_0$ holds as well as
in models where $A$ and $B$ are independent
parameters\refmark\diazlglu. This conclusion is based on
universal scalar masses at the unification scale, and it would be
interesting to see the effect of the relaxation of this hypothesis
(see for example ref.[\pomnir]).


\noindent{\bf ACKNOWLEDGMENTS:}
Useful comments by Howard Baer, Manuel Drees, Tonnis ter Veldhuis,
and Thomas J. Weiler are acknowledged.
This work was supported by the U.S. Department of Energy, grant No.
DE-FG05-85ER-40226, and by the Texas National Research Laboratory
(SSC) Commission, award No. RGFY93-303.

\refout
\endpage
\figout
\end